\documentclass[prd,aps,showpacs,preprintnumbers,amssymb,nofootinbib]{revtex4}

\usepackage{epsf}

\def\e3p{$\eta \rightarrow 3 \pi$}

\begin{document}

\title{%
\hfill{\normalsize\vbox{%
 }}\\
{Isosinglet Scalar Mesons Below 2 GeV and the Scalar Glueball Mass}}

\author{Amir H. Fariborz \footnote[1]{Email:  fariboa@sunyit.edu}}

\affiliation{Department of Mathematics/Science,\\
State University of New York Institute of Technology, Utica, New York 
13504-3050.} 

\vskip 2cm

\date{\today}

\begin{abstract}

A collective treatment of the $I=0$ scalar mesons below 2 GeV [$\sigma
(550)$, $f_0(980)$, $f_0(1370), f_0(1500)$ and $f_0(1710)$] in a
non-linear chiral Lagrangian framework that is constrained by the mass and
the partial decay widths of the $I=1/2,1$ scalars [$\kappa(900),
K_0^*(1430), a_0(980)$ and $a_0(1450)$] is presented.  The sub-structure
of these states in terms of two and four quark components, as well as a
glueball component is explored, and its correlation with the mass of
$f_0(1370)$ is studied.  Consistency with the available experimental data
suggests that the $\sigma(550)$ is dominantly a non-strange four-quark
state, whereas the sub-structure of other $I=0$ states are sensitive to
the input mass of $f_0(1370)$. This investigation estimates the scalar
glueball mass in the range 1.47--1.64 GeV.

\end{abstract}

\pacs{13.75.Lb, 11.15.Pg, 11.80.Et, 12.39.Fe}

\maketitle

\section{Introduction}

Scalar mesons play important roles in low-energy QCD, and
are at the focus of many theoretical and experimental
investigations.  Scalars are important from the theoretical
point of view because they are Higgs bosons of QCD and
induce chiral symmetry breaking, and therefore, are probes
of the QCD vacuum.  Scalars are also important from a
phenomenological point of view, as they are very important
intermediate states in Goldstone boson interactions away
from threshold, where chiral perturbation theory is not
applicable. There are 9 candidates for the lowest-lying
scalar mesons ($ m < 1$ GeV): $f_0(980)$ [$I=0$] and
$a_0(980)$ [$I=1$] which are well established
experimentally \cite{PDG};  $\sigma (560)$ or $f_0(600)$
[$I=0$] with uncertain mass and decay width \cite{PDG}; and
$\kappa(900)$ [$I=1/2$] which is not listed but mentioned
in PDG \cite{PDG}.  The $\kappa(900)$ is
observed in some theoretical models
\cite{BFSS1,vanBeveren,Ishida_kappa}, as well as in some
experimental investigations \cite{E791}.
It is known that a simple $q
{\bar q}$ picture does not explain the properties of these
mesons.  Different theoretical models that go beyond a
simple $q {\bar q}$ picture have been developed, including:
MIT bag model \cite{Jaf}, $K {\bar K}$ molecule \cite{Isg},
unitarized quark model \cite{vanBeveren,Tor}, QCD sum-rules
\cite{Eli}, and chiral Lagrangians
\cite{BFSS1,San,BFSS2,Far,pieta,Mec,LsM,AS,Pelaez}.

The next-to-lowest scalars (1 GeV $< m <$ 2 GeV) are:
$K_0^*(1430)$ [$I=1/2$]; $a_0(1450)$ [$I=1$]; $f_0(1370)$,
$f_0(1500)$, $f_0(1710)$ [$I=0$], and are all listed in
\cite{PDG}. The $f_0(1500)$ is believed to contain a large
glue component and therefore a good candidate for the
lowest scalar glueball state.  These states, are generally
believed to be closer to $q {\bar q}$ objects; however,
some of their properties cannot be explained based on a
pure $q{\bar q}$ structure.

Chiral Lagrangians, provide a powerful framework for
studying the lowest and the next-to-lowest scalar states
probed in different Goldstone boson interactions ($\pi\pi$,
$\pi K$, $\pi\eta$,...) away from threshold
\cite{San,BFSS1,BFSS2,Far,pieta,LsM}. In this approach, a
description of the scattering amplitudes which are, to a
good approximation, both crossing symmetric and unitary is
possible.  To construct scattering amplitudes, all
contributing intermediate resonances up to the energy of
interest are considered, and only tree diagrams (motivated
by large $N_c$ approximation) are taken into account.  In
this way, crossing symmetry is satisfied, but the
constructed amplitudes should be regularized.  
Regularization procedure in turn unitarizes the scattering
amplitude.  By fitting the resulting scattering amplitude
to experimental data, the unknown physical properties
(mass, decay width, ...) of the light scalar mesons can be
extracted.  It is shown in \cite{San} that there is a need
for a $\sigma$ meson with a mass around 550 MeV in order to
describe the experimental data on $\pi\pi$ scattering
amplitude.  Similarly, in \cite{BFSS1} the need for a
$\kappa$ meson with a mass around 900 MeV for describing
the available data on $\pi K$ scattering amplitude is
presented.  Motivated by the evidence for a $\sigma$ and a
$\kappa$, and taking into account the experimentally
well-established scalars, the $f_0(980)$ and the
$a_0(980)$, a possible classification of these states (all
below 1 GeV)  into a lowest-lying scalar meson nonet is
investigated in \cite{BFSS2}.  In this approach, the
non-linear chiral Lagrangian is expressed in terms of this
nonet, and it is shown \cite{BFSS2} that the consistency
with several low energy processes ($\pi \pi$ and $\pi K$
scatterings, and $\Gamma [f_0(980)\to \pi\pi]$)  requires a
scalar mixing angle which is more consistent with a
four-quark assignment for the lowest-lying scalar states.
This model also well describes the experimental
measurements for the $\eta' \to \eta \pi \pi$ decay
\cite{Far}, and estimates the total decay width of
$a_0(980)$ to be around 70 MeV consistent with a recent
experimental confirmation \cite{Teige}.  This model has
also been employed to describe the radiative decays
$\phi\to\pi\eta\gamma$ and $\phi\to\pi\pi\gamma$ in ref.
\cite{Blk_rad}.

In order to study properties of the $a_0(1450)$ and the
$K_0^*(1430)$, in the framework of a non-linear chiral, a
next-to-lowest lying scalar nonet is introduced in
\cite{Mec}.  Similar two-nonet treatment of the scalar
mesons in the context of linear sigma model are
investigated in \cite{LsM,Close2}.  It is shown in
\cite{Mec} that a mixing between the lowest nonet (a
four-quark nonet) and the next-to-lowest nonet (a two-quark
nonet) provides a description of the mass spectrum and the
partial decay widths of the $I=1/2$ and $I=1$ scalars
[$\kappa(900), a_0(980), K_0^*(1430)$ and $a_0(1450)$]. The
case of $I=0$ states is not studied in \cite{Mec} and is
the goal of the present work.  Specifically, we would like
to see if the $I=0$ scalar states below 2 GeV can be
described within a framework that is constrained by the
mixing scenario of ref. \cite{Mec}, and in addition
includes terms that are relevant to $I=0$ states only, such
as mixing with a scalar glueball. The mixing of scalar
glueballs with ${\bar q} q$ scalar mesons has been studied
in the literature \cite{Close,Teshima}.  However, the
importance of mixing between the ${\bar q} q$ and ${\bar q}
{\bar q} q q$ nonets in ref. \cite{Mec} suggests that the
mixing of glueballs with the four-quark nonet should also
be examined.  Therefore, in the present work we investigate
a description of the $I=0$ states in terms of two and four
quark components, and a glueball component. Moreover, we
investigate the sensitivity of the results on the input
data.

After a brief review of the mixing mechanism of ref.
\cite{Mec} in Sec. II, we study the mass spectrum and the
sub-structure of the isosinglet states in Sec. III.  A
description of the two-pseudoscalar decay widths of these
states is investigated in Sec. IV, followed by a summary
and conclusion in Sec.  V.  The basic formulas are listed
in the appendixes.

\section{Mixing Mechanism for I=1/2 and I=1 Scalar States}

In ref. \cite{Mec} the properties of the $I=1/2$ and $I=1$ scalar mesons,
$\kappa(900), K_0^*(1430), a_0(980)$ and $a_0(1450)$, in a non-linear
chiral Lagrangian framework is studied in detail. In this approach, a
${\bar q} {\bar q} q q$ nonet $N$ mixes with a ${\bar q} q$ nonet $N'$ and
provides a description of the mass spectrum and decay widths of these
scalars.  This mixing  provides an explanation for some 
unexpected
properties of the $K_0^*(1430)$ and the $a_0(1450)$, which are generally
believed to be good candidates for a ${\bar q} q$ nonet \cite{PDG}, but
some of their properties do not quite follow this scenario. For example,
in a $q {\bar q}$ nonet, isotriplet is expected to be lighter than the
isodoublet, but for these two states \cite{PDG}:
\begin{equation}
m \left[ a_0(1450) \right] = 1474 \pm 19 \hskip .2cm {\rm MeV} > m 
\left[ K_0^*(1430) \right] =
1412 \pm 6  \hskip .2cm {\rm MeV}
\label{a0k0_mass_exp}
\end{equation}
Also their decay ratios given in PDG \cite{PDG} do not closely follow a 
pattern 
expected from an SU(3) symmetry (given in parenthesis):
\begin{equation}
{ {\Gamma \left[ a_0^{total} \right]}\over
{\Gamma\left[K_0^*\rightarrow\pi K\right] }} = 0.92 \pm 0.12 
\hskip .2cm (1.51),
\hskip 0.2cm
{ {\Gamma \left[ a_0\rightarrow K{\bar K} \right]}\over
{\Gamma\left[a_0\rightarrow\pi \eta\right] }} = 0.88 \pm 0.23 
\hskip .2cm  (0.55),
\hskip 0.2cm
{ {\Gamma \left[ a_0\rightarrow \pi\eta' \right]}\over
{\Gamma\left[a_0\rightarrow\pi \eta\right] }} = 0.35\pm0.16 \hskip 
.2cm  (0.16)
\label{a0k0_decay_exp}
\end{equation}
These properties of the $K_0^*(1430)$ and the $a_0(1450)$ are naturally  
explained by the mixing mechanism of ref. \cite{Mec}.   
The general mass terms contributing to  the $I=1/2$ and the $I=1$ states 
can be written as
\begin{equation}
{\cal L}_{mass}^{I=1/2,1} = 
- a {\rm Tr}(NN) - b {\rm Tr}(NN{\cal M}) 
- a' {\rm Tr}(N'N') - b' {\rm Tr}(N'N'{\cal M}) 
\label{L_mass_I1}
\end{equation}
where ${\cal M} = {\rm diag} (1,1,x)$ with $x$ being the ratio of
the strange to non-strange quark masses, and $a,b,a'$ and $b'$ are 
unknown parameters fixed by the ``bare'' masses:
\begin{equation}
\begin{array}{cclcccl}
m^2 [a_0] &=& 2 (a + b) & &  m^2 [a'_0] & = & 2 (a' + b')\\
m^2 [K_0] &=& 2a + (1+x) b & &  m^2 [K'_0] & = & 2 a' + (1+x) b'
\end{array}
\end{equation}
where the subscript ``0'' denotes the ``bare'' states (i.e. before the 
mixing 
between $N$ and $N'$ is taken into account).    Therefore, as $N$ is a 
four-quark nonet and $N'$ a two-quark nonet, we expect:
\begin{equation}
m^2 [K_0] <  m^2 [a_0] \le m^2 [a'_0] < m^2 [K'_0] 
\label{mass_order}
\end{equation}
Introducing a simple mixing 
\begin{equation}
{\cal L}_{mix}^{I=1/2,1} = 
-\gamma {\rm Tr} \left( N N' \right) 
\label{L_mix_I1}
\end{equation}
it is shown in \cite{Mec} that  for
$0.51 < \gamma < 0.62 \hskip .2cm{\rm GeV^2}$, it is possible to 
recover the physical masses such that the ``bare'' masses have the 
expected ordering of (\ref{mass_order}).
Therefore in this mechanism, the ``bare'' isotriplet states split more 
than the isodoublets, and 
consequently, the
physical isovector state $a_0(1450)$ becomes heavier than the  
isodoublet state $K_0^*(1430)$ as observed in (\ref{a0k0_mass_exp}).   The 
light isovector and 
isodoublet states are the $a_0(980)$ and the
$\kappa(900)$.    With the physical masses $m[a_0(980)]=0.9835$ GeV, 
$m [\kappa(900)] =0.875$ GeV, $m[a_0(1450)]=1.455$ GeV and 
$m[K_0(1430)]=1.435$ GeV, the best values of $\gamma$ and the ``bare''
masses are found in \cite{Mec}
\begin{equation}
m_{a_0}=m_{a'_0}=1.24 \hskip .1cm{\rm GeV},\hskip .2cm 
m_{K_0}=1.06 \hskip .1cm{\rm GeV}, \hskip .2cm 
m_{K'_0}=1.31 \hskip .1cm{\rm GeV}, \hskip .2cm 
\gamma = 0.58 \hskip .1cm {\rm GeV^2}
\label{a0k0_m_bare}
\label{bare_masses}
\end{equation}

The decay ratios (\ref{a0k0_decay_exp}) are also investigated in 
\cite{Mec} in 
which the scalar-pseudoscalar-pseudoscalar interaction part relevant to 
the $I=1/2$ and $I=1$ scalar states is given as
\begin{equation}
{\cal L}_{int.}^{I=1/2,1} =
A \epsilon^{abc}\epsilon_{def}
N_a^d\partial_\mu\phi^e_b\partial_\mu\phi^f_c
+ C {\rm Tr} (N \partial_\mu \phi) {\rm Tr} (\partial_\mu\phi)
+ A' \epsilon^{abc}\epsilon_{def}
{N'}_a^d\partial_\mu\phi^e_b\partial_\mu\phi^f_c
+ C' {\rm Tr} (N' \partial_\mu \phi) {\rm Tr} (\partial_\mu\phi)
\label{L_int_I1}
\end{equation}
where $A, C, A'$ and $C'$ are unknown parameters fixed by decay properties 
of the scalars, and $\phi$ is the conventional pseudoscalar meson nonet.
It is shown in \cite{Mec} that with parameters
\begin{equation}
A=1.19\pm 0.16 \hskip .1cm {\rm GeV}^{-1}, \hskip 0.2cm 
A'=-3.37\pm 0.16 \hskip .1cm  {\rm GeV}^{-1}, \hskip 0.2cm 
C=1.05\pm 0.49 \hskip .1cm  {\rm GeV}^{-1}, \hskip 0.2cm 
C'=-6.87\pm 0.50 \hskip .1cm  {\rm GeV}^{-1}, \hskip 0.2cm 
\label{AC_values}	
\end{equation}
a reasonable agreement on the decay ratios of the $K_0^*(1430)$ and the 
$a_0(1450)$ consistent with (\ref{a0k0_decay_exp}), as well as the 
expected decay 
widths of the $a_0(980)$ and the $\kappa(900)$ can be obtained.
In next section, we include the Lagrangians (\ref{L_mass_I1}) and 
(\ref{L_int_I1}) together with parameters (\ref{a0k0_m_bare}) and 
(\ref{AC_values}) as part of the mass and interaction Lagrangian of 
the $I=0$ scalar states.

\section{Isosinglet States}

The general mass terms for nonets $N$ and $N'$, and a scalar glueball $G$ 
can be 
written as:
\begin{equation}
{\cal L}_{mass}^{I=0}  = {\cal L}_{mass}^{I=1/2,1} 
- c {\rm Tr}(N){\rm Tr}(N) 
- d {\rm Tr}(N) {\rm Tr}(N{\cal M}) 
- c' {\rm Tr}(N'){\rm Tr}(N') 
- d' {\rm Tr}(N') {\rm Tr}(N'{\cal M}) 
- g G^2
\label{L_mass_I0}
\end{equation}
The a priori unknown parameters $c$ and $d$ induce ``internal'' mixing 
between the two $I=0$ flavor combinations [$(N_1^1 + N_2^2)/\sqrt{2}$ and 
$N_3^3$] of nonet $N$. Similarly,  $c'$ and $d'$ play the same 
role in nonet $N'$.  
Parameters $c,d,c'$ and $d'$ do not contribute to 
the mass spectrum of the $I=1/2$ and $I=1$ states.	
The last term represents the glueball mass term.  
The term ${\cal 
L}_{mass}^{I=1/2,1}$ is imported from Eq. (\ref{L_mass_I1}) together with 
its parameters from Eq. 
(\ref{a0k0_m_bare}).

\begin{figure}[htbp]
\begin{center}
\epsfxsize = 10cm
\epsfbox{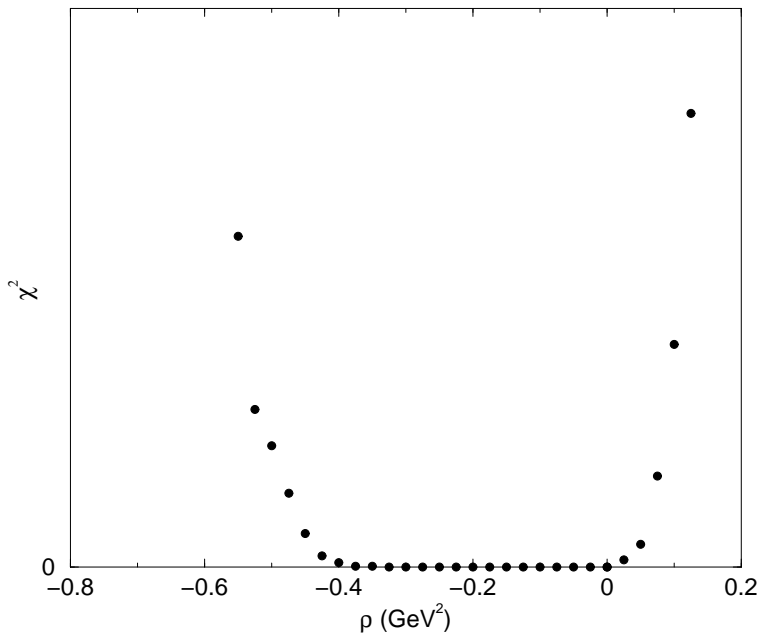}
\end{center}
\caption{$\chi^2$ vs $\rho$}
\label{F_chi2_vs_rho}
\end{figure}

The mixing between $N$ and 
$N'$, and the mixing of these two nonets with the  scalar glueball $G$ can 
be 
written as
\begin{equation}
{\cal L}_{mix}^{I=0} = 
{\cal L}_{mix}^{I=1/2,1}
- \rho {\rm Tr} (N) {\rm Tr} (N')
- e G {\rm Tr} \left( N \right)
- f G {\rm Tr} \left( N' \right)
\label{L_mix_I0}
\end{equation}
where the first term is given in 
(\ref{L_mix_I1}) with 
$\gamma$ from (\ref{a0k0_m_bare}).   The second term does not 
contribute to the $I=1/2,1$ mixing, and in  special limit of 
$\rho\to-\gamma$:
\begin{equation}
- \gamma  {\rm Tr} (N N')
- \rho {\rm Tr} (N) {\rm Tr} (N')
= \gamma \epsilon^{abc} \epsilon_{ade} N^d_b {N'}^e_c
\label{mix_OZI}
\end{equation} 
This particular mixing is more consistent with the OZI rule than 
the individual $\gamma$ and $\rho$ terms and is studied in \cite{Teshima}.    
Here we do not restrict the mixing to this particular combination, and 
instead, examine a range of $\rho$ values.
Terms with unknown couplings $e$ and $f$
describe mixing with the scalar glueball $G$.    
As a result, the five isosinglets 
below
2 GeV, become a mixture of five different flavor combinations, and their 
masses can be organized as
\begin{equation}
{\cal L}_{mass}^{I=0} + {\cal L}_{mix}^{I=0} = - {1\over 2} {\tilde {\bf 
F}}_0 {\bf M}^2 {\bf 
F}_0 = 
- {1\over 2} {\tilde {\bf F}} {\bf M}_{diag.}^2 {\bf F}  
\label{L_mass_and_mix}
\end{equation}
with
\begin{equation}
{\bf F}_0 =
       \begin{array}{l}
                         \left(
                         \begin{array}{c}
                              N_3^3 \\
                             (N_1^1 + N_2^2)/\sqrt{2}\\
                              {N'}_3^3 \\
                             ({N'}_1^1 + {N'}_2^2)/\sqrt{2}\\
                              G  
                         \end{array}
                         \right)
                   
       \end{array}   
=
       \begin{array}{l}
                         \left(
                         \begin{array}{c}
      {\bar u}{\bar d} u d\\
     ({\bar s}{\bar d} d s + {\bar s}{\bar u} u s) /\sqrt{2}\\
                              {\bar s}s\\
                             ({\bar u} u + {\bar d} d)/\sqrt{2}\\
                              G  
                         \end{array}
                         \right)
                   
       \end{array}   
=
       \begin{array}{l}
                         \left(
                         \begin{array}{l}
                             f_0^{NS}\\
                             f_0^S \\
                             {f'}_0^S\\
                             {f'}_0^{NS} \\
                              G  
                         \end{array}
                         \right)
       \end{array}   
\label{SNS_def}
\end{equation}
where the superscript $NS$ and $S$ respectively represent the non-strange 
and strange combinations.  ${\bf F}$ contains the physical fields
\begin{equation}
{\bf F}  =
       \begin{array}{c}
                        \left( \begin{array}{c}
                                \sigma(550)\\
                                f_0(980)\\
                                f_0(1370)\\
                                f_0(1500)\\
                                f_0(1710)
                                \end{array}
                         \right)  
= K^{-1} {\bf F}_0
       \end{array} \label{K_def} 
\end{equation} 
where $K^{-1}$ is the transformation matrix. The mass 
squared
matrix is 
\begin{equation} 
{\bf M}^2 = \left[
\begin{array}{ccccc} 
2 m_{K_0}^2 - m_{a_0}^2 + 2 (c + d x)
& \sqrt{2}[2c + (1+x)d] & \gamma + \rho& \sqrt{2} \rho & e
\\ 
\sqrt{2}[2c + (1+x)d] & m_{a_0}^2 + 4 (c + d) & \sqrt{2}
\rho & \gamma + 2 \rho& \sqrt{2} e 
\\ 
\gamma + \rho&
\sqrt{2} \rho & 2 m_{K'_0}^2 - m_{a'_0}^2 + 2 (c' + d' x) &
\sqrt{2}[2c' + (1+x)d'] & f 
\\ 
\sqrt{2} \rho & \gamma + 2
\rho& \sqrt{2}[2c' + (1+x)d'] & m_{a'_0}^2 + 4 (c' + d') &
\sqrt{2} f 
\\ 
e & \sqrt{2} e & f & \sqrt{2} f & 2g
\end{array} 
\right] 
\label{mass_matrix} 
\end{equation} 
in which the value of the unmixed $I=1/2,1$ masses, and the
mixing parameter $\gamma$ are substituted in from
(\ref{bare_masses}).  We search for the unknown parameters
$c,c',d,d',e,f,g$ and $\rho$ in (\ref{mass_matrix}) by
fitting its eigenvalues to the mass of the physical states.  
We take $m_\sigma = 550 \pm 50$ MeV expected from chiral
Lagrangian treatment of the $\pi\pi$ scattering in
\cite{San}, as well as the following experimental values
from PDG \cite{PDG}: 
\begin{equation} 
\begin{array}{r c l}
m [f_0(980)] &=& 980 \pm 10 \, {\rm MeV} \\ m [f_0(1370)]
&=& 1200 \rightarrow 1500 \, {\rm MeV} \\ m [f_0(1500)] &=&
1507 \pm 5 \, {\rm MeV} \\ m [f_0(1710)] &=& 1713 \pm 6 \,
{\rm MeV} \label{exp_masses} 
\end{array} 
\end{equation} 
The largest experimental uncertainty is on the mass of
$f_0(1370)$, and for our initial study we take its central
value of 1350 MeV.  The sensitivity of the results on the
mass of $f_0(1370)$ turns out to be the main source of
error and will be discussed later. Figure
\ref{F_chi2_vs_rho} shows the dependency of $\chi^2$ on
parameter $\rho$.  We see that for $-0.4 \,{\rm GeV^2} < 
\rho < 0$ the $\chi^2$ is very small, but significantly
increases outside this interval.  For five values of $\rho$
the result of fits are given in table \ref{T_RHO}, in which
the sensitivity of the fitted parameters on $\rho$ can be
seen.  As far as fitting the isosinglet masses is
concerned, all chosen values of $\rho$ in table \ref{T_RHO}
give more or less the same description.  Parameters $d$ and
$d'$ induce SU(3)  symmetry breaking and are at least an
order of magnitude smaller than $c$ and $c'$. The fitted
parameters determine the rotation matrix (\ref{K_def})
which in turn probes the quark substructure of the scalars.  
For $\rho=0$ the rotation matrix is given in
(\ref{K_inverse_rho0}) in which the errors reflect the
experimental uncertainties in (\ref{exp_masses}).  For
other values of $\rho$ in table \ref{T_RHO}, the
corresponding rotation matrices are given in Appendix A.
The overall results show that the sub-structure of
$\sigma(550)$ is not sensitive to $\rho$ and is dominantly
a non-strange four quark combination ${\bar u}{\bar d} u
d$, consistent with the investigation of ref.  
\cite{BFSS2}. The sub-structure of the $f_0(980)$ is more
sensitive to $\rho$; for $\rho < -0.2 \, {\rm GeV}^2$ it 
has a
dominant ${\bar s}s$ component, whereas for $\rho \ge 
-0.2\,
{\rm GeV}^2$ the non-strange $({\bar u} u + {\bar d}
d)/\sqrt{2}$ component dominates.  The later case has some
support in QCD sum-rules \cite{Eli}. The $f_0(1370)$ has
substantial two-quark components with some glueball
admixture, and the $f_0(1500)$ and $f_0(1710)$ have large
glueball components.  For each value of $\rho$, the
corresponding glueball mass is also given in table
\ref{T_RHO}, showing a variation in the range 1.47-1.60
GeV.  As $\rho \to -\gamma$ the result indicates that the
glueball component of the $f_0(1500)$ significantly
increases, but in this limit the $\chi^2$ is relatively
large, and therefore, the result is not as accurate as the
cases given in table \ref{T_RHO}.  For $\rho=-0.5 \, {\rm
GeV}^2$ (which is close to the limit $\rho \to -\gamma$)  
the rotation matrix is given in Eq.
(\ref{mass_matrix_rhom5}), and is consistent with the 
result of ref. \cite{Teshima}.

\begin{table}[htbp]
\begin{center}
\begin{tabular}{cccccc}
\hline
\hline
Fitted Parameters & $\rho=-0.5 {\rm GeV}^2$   & $\rho=-0.4  {\rm GeV}^2$  
& $\rho = -0.2 {\rm GeV}^2$ 
& $\rho=0$ &  $\rho=0.075 {\rm GeV}^2$  
\\
\hline 
$c$ (GeV$^2$)     & 
0.141  & 0.181   & 
0.281  &  
$0.157 \pm 0.009$  &  
0.201
 \\
$d$ (GeV$^2$)     & 
$-0.00494$ & $-0.0106$  & 
$-0.0175$ & 
$-0.00779\pm 0.00187$  & 
$-0.00840$
 \\
$c'$(GeV$^2$)     & 
$-0.0389$    & $-0.0475$    & 
$-0.0356 $ & 
$-0.0128 \pm 0.0052$  &
$-0.00184$ 
  \\
$d'$ (GeV$^2$)    & 
$-0.00138$  &  $-0.00407$   & 
$-0.00452$  & 
$-0.00927\pm 0.00116$ & 
$-0.00825$ 
  \\
$e$ (GeV$^2$)     & 
$-0.169$    &  $-0.165$  & 
$-0.242$ & 
$-0.323 \pm 0.018$    & 
$-0.248$
\\
$f$ (GeV$^2$)     & 
0.0941     & 0.234       
& 0.323  & 
$0.115 \pm 0.0568$   &  
0.0874
\\
$g$ (GeV$^2$)     & 
1.211     &  1.249   & 
1.080  &     
$1.272\pm 0.021$     & 
1.159
\\
$m_G$ (GeV)       & 
1.556     & 1.581    & 
1.470 & 
$1.595 \pm 0.013$       & 
1.523
\\
$\chi^2$          & 
0.0740    &  $\approx 0$      
& $\approx 0$
& $ \approx 0$            
& 0.0932  
\\
\hline
\hline
\end{tabular}
\end{center}
\caption[]{Best numerical values for the unknown parameters in  
Lagrangian (\ref{L_mass_and_mix}), with $m[f_0(1370)]=1.35$ GeV and 
several 
values of $\rho$.   
For $\rho=0$, the errors on the fitted 
parameters are also given, and reflect the experimental uncertainties in 
(\ref{exp_masses}). 
} 
\label{T_RHO} 
\end{table} 

\begin{equation}
K^{-1}_{(\rho=0)}  =
\left[ \begin{array}{ccccc}
0.853 \pm 0.015 &  -0.011 \pm 0.024  &  -0.476 \pm 0.018 & 
-0.148 \pm 
0.024 &
0.156 \pm 0.017\\
0.278 \pm 0.040 &  -0.497 \pm 0.009  &  0.211  \pm 0.019 & 0.774 \pm 0.013   
& 
-0.181 \pm 0.035\\
0.394 \pm 0.020 &  0.311 \pm 0.026   &  0.817 \pm 0.025  & -0.102 \pm 
0.022 & 0.266 
\pm 0.048 \\
-0.125 \pm 0.021& 0.550 \pm 0.024    &  -0.247 \pm 0.055 &  0.588 \pm 
0.015 & 
0.525 \pm 0.020 \\
-0.157 \pm 0.023  & -0.595 \pm 0.017 &  0.032 \pm 0.019  &  -0.153 \pm 
0.017   
&  
0.773 \pm 0.016
\end{array}
\right]
\label{K_inverse_rho0}
\end{equation}

\begin{figure}[htbp]
\begin{center}
\epsfxsize = 10cm
\epsfbox{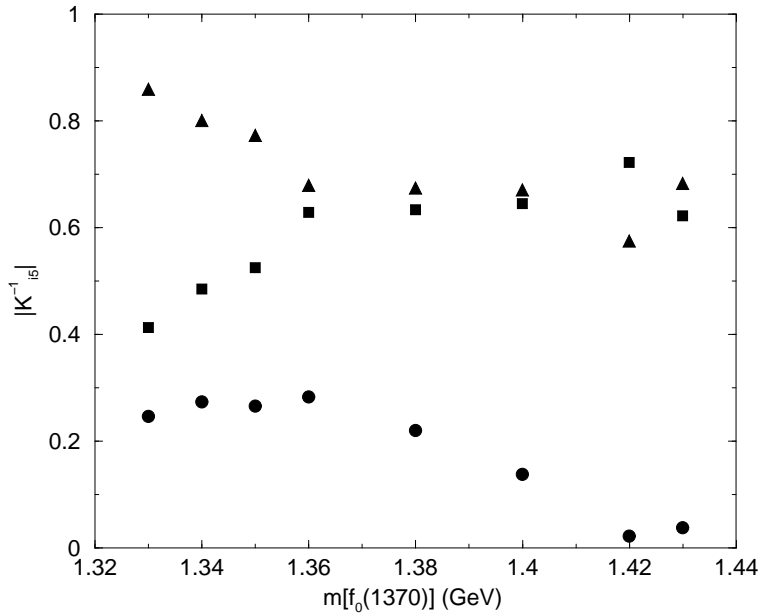}
\end{center}
\caption
{The dependence on the $m[f_0(1370)]$ of the magnitude of the glueball 
component of $f_0(1370)$ (circles),  $f_0(1500)$ 
(squares), and  $f_0(1710)$ (triangles).}
\label{F_K_vs_MF0}
\end{figure}

The fits are sensitive to the input mass of $f_0(1370)$
which has the largest experimental uncertainty in
(\ref{exp_masses}). We find that very good $\chi^2$ fits
can be obtained for $1.31 {\rm GeV} \le m[f_0(1370)] \le
1.45 {\rm GeV}$.  Outside this interval, however, the
goodness of the fits significantly reduces.  This
observation further restricts the $m[f_0(1370)]$ in
(\ref{exp_masses}).  For several values of $m[f_0(1370)]$
the fitted parameters are given in table \ref{T_MF0}, and
the corresponding rotation matrices are given in Appendix
A. The resulting glueball mass is in the range
$ 
1.54 \, {\rm GeV} \le m_G
\le 1.61 \, {\rm GeV} 
$. 
For this range of $m[f_0(1370)]$, the sensitivity of the
glueball content of the $f_0(1370)$, $f_0(1500)$ and
$f_0(1710)$ on the mass of $f_0(1370)$ are examined in
figure \ref{F_K_vs_MF0} and show a strong dependence.  For
example, we see that below the central value of
$m[f_0(1370)]=1.35$ GeV, the glueball component of the
$f_0(1710)$ dominates those of the $f_0(1500)$ and
$f_0(1370)$.  However, above $m[f_0(1370)] \approx 1.4$
GeV, the $f_0(1500)$ acquires the largest glue component
followed by the components of the $f_0(1710)$ and
$f_0(1370)$.  The order again changes around $m[f_0(1370)]
\approx 1.42$ GeV.  Of course this periodic-like behavior
is not surprising as the rotation matrix, which depends on
ten mixing angles, goes through periodic variations as we
tune $m[f_0(1370)]$.  Therefore, due to this sensitivity on
the mass of $f_0(1370)$, a precises extraction of the
glueball content of these states requires an accurate
knowledge of the $m[f_0(1370)]$.

The correlation between the mixing parameter $\rho$ and the
input mass for the $f_0(1370)$ are also examined.  
Although the texture of the rotation matrix varies with
these two parameters, the glueball mass remains more 
or less within the same intervals obtained by 
uncorrelated variations of $\rho$ and $m[f_0(1370)]$ (see 
above).   The overall numerical work shows
\begin{equation}
1.47 \, {\rm GeV} \le m_G
\le 1.64 \, {\rm GeV} \label{m_G} 
\end{equation} 
in agreement with the
lattice QCD estimates \cite{Lattice}. 
There are values of $\rho$
and $m[f_0(1370)]$, for which the glueball masses are as
low as 1.32 GeV. However, for these cases the $f_0(980)$
acquires a large glueball component which is not consistent
with either the molecule or the four-quark description of
this state.

\begin{table}[htbp]
\begin{center}
\begin{tabular}{c|cccc}
\hline
\hline
Fitted  & & $m[f_0(1370)]$ (GeV)& & 
\\
Parameters        & 
1.34              & 
1.36              & 
1.38              & 
1.40  
\\
\hline 
$c$ (GeV$^2$)     & 
0.143             & 
0.190             &  
0.193             &
0.197
 \\
$d$ (GeV$^2$)     & 
$-0.00692$        & 
$-0.00921$        & 
$-0.00968$         & 
$-0.0104$
 \\
$c'$(GeV$^2$)     & 
$-0.00773$        & 
$-0.0161$         & 
$-0.0167$         & 
$-0.0215$ 
  \\
$d'$ (GeV$^2$)    & 
$-0.0104$         &  
$-0.00908$        & 
$-0.00699$        & 
$-0.00440$ 
  \\
$e$ (GeV$^2$)     & 
$-0.318$          &  
$-0.319$          &
$-0.337$          & 
$-0.344$
\\
$f$ (GeV$^2$)     & 
0.101             & 
0.125             & 
$0.167$           &  
0.188
\\
$g$ (GeV$^2$)     & 
1.291             &  
1.221             &     
$1.204$           & 
1.194
\\
$m_G$ (GeV)       & 
1.607             & 
1.563             &  
1.552             & 
1.546
\\
\hline
\hline
\end{tabular}
\end{center}
\caption[]{Best numerical values for the unknown parameters in  
Lagrangian (\ref{L_mass_and_mix}), with $\rho=0$  and several values of 
the $m[f_0(1370)]$.
} 
\label{T_MF0} 
\end{table}

It is important to note that the coupling of the glueball
to both nonets $N$ and $N'$ should be taken into account.  
This is examined in table \ref{T_GN} for $\rho=0$ and
$m[f_0(1370)]$=1.35 GeV.  The first column corresponds to
the general case where the coupling of the glueball to both
nonets $N$ and $N'$ is taken into account. We see that
scalar glueball coupling to $N$ is larger than its coupling
to $N'$.  In a recent work \cite{Teshima} a similar
analysis is given in which the coupling to the lowest-lying
nonet is neglected.  To illustrate the importance of the
glueball coupling to both nonets, in the second and third
columns of Table \ref{T_GN} the glueball coupling to $N$
and $N'$ is respectively suppressed.  We see that the
result is sensitive to the glueball coupling to $N$ and
$N'$.  For example, when $e=0$ the glueball mass
significantly increases to 1.45 GeV, whereas when $f=0$ the
glueball mass decreases to 1.22 GeV.  (We see that the
result seems to be more sensitive to $e$.)  Other fitted
parameters in Table \ref{T_GN} also show a similar
sensitivity to the suppression of the couplings $e$ and
$f$.  Therefore, it is important to consider a general
treatment in which both couplings are taken into account.

\begin{table}[htbp]
\begin{center}
\begin{tabular}{cccc}
\hline
\hline
Fitted Parameters & Fit 1   & Fit 2    
& Fit 3\\
\hline 
$c$ (GeV$^2$) & $0.16$    & $0.081 $   &  $0.21$   \\
$d$ (GeV$^2$) & $-0.0078$ & $-0.0067 $ &  $-0.0098$ \\
$c'$(GeV$^2$) & $-0.013 $ & $-0.027$   & $-0.013 $    \\
$d'$ (GeV$^2$)& $-0.0093$ & $-0.0058$  & $-0.013$       \\
$e$ (GeV$^2$) & $-0.32$   & $0$        & $-0.25$      \\
$f$ (GeV$^2$) & $0.12$    & $0.13$     &  0     \\
$g$ (GeV$^2$) & $1.27$    & $1.45$     & $1.22$      \\
\hline
\hline
\end{tabular}
\end{center}
\caption[]{Numerical values for the unknown parameters in  
Lagrangian (\ref{L_mass_and_mix}).
The first column corresponds to the case where the scalar glueball couples 
to 
both the light scalar meson nonet $N$ as well as to the heavy scalar 
meson nonet $N'$.    The second and third columns represent cases where
the scalar glueball only couples to the nonet $N'$, and to the nonet $N$, 
respectively.} 
\label{T_GN} 
\end{table}

\section{Interaction Lagrangian}

To further examine the present model for the scalar mesons, we need to
investigate the interaction Lagrangian for these states and study their
partial decay widths to several two-pseudoscalar channels.  We saw in
previous section that the experimental input mass of $f_0(1370)$ is the
main source of uncertainty on the rotation matrix which in turn affects
the interaction Lagrangian and decay width calculations.  Therefore,
without an accurate knowledge of $m[f_0(1370)]$ we can only give a
qualitative description of the decay properties. Moreover, although
various decay ratios are recently reported by the WA102 collaboration
\cite{WA102} (see table \ref{T_BBpDDpEF_fit}), the experimental status of
the individual decay widths is not quite clear.  In PDG \cite{PDG}
some of these decay widths are given, even though they are not used in any
averaging.  Nevertheless, in this section we present the interaction
Lagrangian and give a preliminary study of several decay ratios.  This
will provide a basis in which the upcoming experimental data on the decay
widths/ratios of the scalar mesons can be analyzed.

The scalar-pseudoscalar-pseudoscalar interaction takes the general form:
\begin{eqnarray}
{\cal L}_{int.}^{I=0} &=&
{\cal L}_{int.}^{I=1/2,1} +
B {\rm Tr} \left( N \right) {\rm Tr} \left({\partial_\mu}\phi
{\partial_\mu}\phi \right) 
+ D {\rm Tr} \left( N \right) {\rm Tr}
\left({\partial_\mu}\phi \right)  {\rm Tr} \left( {\partial_\mu}\phi
\right)
+ B' {\rm Tr} \left( N' \right) {\rm Tr} \left({\partial_\mu}\phi
{\partial_\mu}\phi \right) 
\nonumber\\
&& + D' {\rm Tr} \left( N' \right) {\rm Tr}
\left({\partial_\mu}\phi \right)  {\rm Tr} \left( {\partial_\mu}\phi
\right)
+ E G  {\rm Tr} \left({\partial_\mu}\phi
{\partial_\mu}\phi \right) 
+ F G  {\rm Tr}
\left({\partial_\mu}\phi \right)  {\rm Tr} \left( {\partial_\mu}\phi
\right)
\label{L_int_I0}
\end{eqnarray}
where $B$ and $D$ are unknown coupling constants describing the 
coupling of the four-quark nonet $N$ to the pseudoscalars.  Similarly, 
$B'$ and $D'$  are couplings of $N'$ to the pseudoscalars. 
$E$ and $F$
describe the coupling of a scalar glueball to the pseudoscalar mesons, and    
${\cal L}_{int.}^{I=1/2,1}$ is the interaction Lagrangian for $I=1/2,1$ 
states given in (\ref{L_int_I1}) together with parameters in 
(\ref{AC_values}).
The 
pseudoscalar part of the Lagrangian can be found in ref. \cite{BFSS1}.
The interaction Lagrangian (\ref{L_int_I0}) can be rewritten as:
\begin{eqnarray}
- {\cal L}_{int} &=&
{1\over \sqrt{2}} \gamma_{\pi\pi}^i \, {\bf F}_i \partial_\mu \pi . 
\partial_\mu 
\pi 
+ {1\over \sqrt{2}} \gamma_{KK}^i \, {\bf F}_i \partial_\mu {\bar K}  
\partial_\mu K
\nonumber \\
&& + \gamma_{\eta\eta}^i \, {\bf F}_i \partial_\mu \eta  
\partial_\mu \eta
+ \gamma_{\eta\eta'}^i \, {\bf F}_i \partial_\mu \eta  
\partial_\mu \eta'
+ \gamma_{\eta'\eta'}^i \, {\bf F}_i \partial_\mu \eta'  
\partial_\mu \eta'
\label{L_int_I0_b}
\end{eqnarray}
where $\gamma^i_{ss'}$ is the coupling of the $i$-th scalar [see Eq. 
(\ref{K_def})] to 
pseudoscalars $s$ and $s'$, and is given by
\begin{equation}
\gamma^i_{ss'} = \sum_j \left( \gamma_{ss'} K\right)_{ji} 
\end{equation}
with  $K$ defined in  ({\ref{K_def}) and 
\begin{equation}
\gamma_{ss'} = \left[
\begin{array}{ccccc}
\gamma^{NS}_{ss'} &                &                &       &          \\
                  & \gamma^S_{ss'} &                &       &          \\
                  &                &\gamma'^S_{ss'} &       &          \\
                  &                &                &\gamma'^{NS}_{ss'}&\\
                  &                &                &       & 
\gamma^G_{ss'}
\end{array}
\right]
\end{equation}           
in which the diagonal elements are the couplings of the 
pseudoscalars $s$ and $s'$ to the $f_0^{NS}, f_0^S, {f'}_0^S, {f'}_0^{NS}$
[defined in (\ref{SNS_def})] and the scalar glueball $G$, respectively.
The diagonal elements for all decay channels $ss'$ are listed in the 
Appendix B.

To determine the unknown couplings we need to fit the prediction of this
Lagrangian to experimental data.  Here we use the estimates of the decay
ratios by the WA102 collaboration \cite{WA102} in table 
\ref{T_BBpDDpEF_fit}.   
We should note, however, that the decay ratios alone are not sufficient to 
determine the
free parameters and need to be supplemented by more data
such as the individual partial decay widths, or the total decay widths 
\cite{PDG}:
\begin{equation}
\begin{array}{r c l}
\Gamma^{total} [f_0(980)] &=& 40 \rightarrow  100 \hskip .2cm {\rm MeV} 
\\
\Gamma^{total} [f_0(1370)] &=& 200 \rightarrow  500 \hskip .2cm {\rm MeV} 
\\
\Gamma^{total} [f_0(1500)] &=& 109 \pm 7 \hskip .2cm {\rm 
MeV} 
\\
\Gamma^{total} [f_0(1710)] &=& 125 \pm  10 \hskip .2cm {\rm MeV} 
\label{masses}
\end{array}
\end{equation}

For example, with $\rho=0$ and $m[f_0(1370)] = 1.35$ GeV, a numerical
study of the decay ratios is given in table \ref{T_BBpDDpEF_fit}, together
with the predicted decay widths in table \ref{T_partial_widths}. The
result is compared with estimates extracted from other works, and shows an
overall qualitative agreement, even though some of the predicted decay
widths such as $\Gamma[\sigma (550)\to \pi\pi]$
 or $\Gamma[f_0(1500)  \to
\pi\pi]$ do not quite agree with other investigations
\footnote{Note that the
$\Gamma[\sigma \to \pi\pi]$, which is proportional to the
$\gamma_{\sigma\pi\pi}^2$, is not the same as the Breit-Wigner
width that appears in the denominator of the $\sigma$
propagator  \cite{San}.  This deviation from a Breit-Wigner shape, is also 
a characteristic of the $\kappa
(900)$ meson \cite{BFSS2}.}.  However we should
again note that the current experimental status of the decay widths is not
quite established, and that together with the large uncertainty on the
mass of the $f_0(1370)$ are the main sources of error in estimates given
in tables \ref{T_BBpDDpEF_fit} and \ref{T_partial_widths}. Certainly more
work is needed to investigate the parameters of the interaction Lagrangian
in more detail, and study its correlation with the mass of $f_0(1370)$. We
 postpone this goal for future works.

\begin{table}[htbp]
\begin{center}
\begin{tabular}{lcc}
\hline
\hline
Fitted Parameters & & \\
\hline 
$B$ (GeV$^{-2}$)     & $ -1.0 \to -0.7 $  & \\
$B'$ (GeV$^{-2}$)    & $ 0.8 \to 1.0 $  & \\
$D$(GeV$^{-2}$)      & $ -0.1 \to 2.5 $ & \\  
$D'$ (GeV$^{-2}$)    & $-3.5 \to 0.5 $ & \\
$E$ (GeV$^{-2}$)     & $-1.6 \to -0.3 $ & \\
$F$ (GeV$^{-2}$)     & $-2.2 \to 2.4 $ & \\
\hline
Decay ratios &  This fit & WA102 Collaboration \\
\hline
$
 { {\Gamma[f_0(1370)\rightarrow \pi\pi]} \over 
   {\Gamma[f_0(1370)\rightarrow K {\bar K}]}
 }
$ 
&    
$1.4 \to 6.5$ 
&
$2.17 \pm 0.9$
\\
$
 { {\Gamma[f_0(1370)\rightarrow \eta\eta]} \over 
   {\Gamma[f_0(1370)\rightarrow K {\bar K}]}
 }
$ 
&
$ < 0.23 $ 
&
$ 0.35 \pm 0.21 $
\\
$
 { {\Gamma[f_0(1500)\rightarrow  K {\bar K}]} \over
   {\Gamma[f_0(1500)\rightarrow \pi\pi }
 }
$
&
$ < 0.16  $ 
&
$ 0.32 \pm 0.07 $
\\
$
 { {\Gamma[f_0(1500)\rightarrow \pi\pi]} \over
   {\Gamma[f_0(1500)\rightarrow \eta\eta }
 }
$
&
$ > 5.3  $ 
&
$ 5.5 \pm 0.84 $
\\
$
 { {\Gamma[f_0(1500)\rightarrow \eta\eta']} \over
   {\Gamma[f_0(1500)\rightarrow \eta\eta }
 }
$
&
$ 0.3 \to 1.7 $ 
&
$ 0.52 \pm 0.16 $
\\
$
 { {\Gamma[f_0(1710)\rightarrow \pi\pi]} \over
   {\Gamma[f_0(1710)\rightarrow K {\bar K} }
 }
$
&
$ 0.19 \to 2.6 $  
&
$ 0.20\pm 0.03 $
\\
$
 { {\Gamma[f_0(1710)\rightarrow \eta\eta]} \over
   {\Gamma[f_0(1710)\rightarrow K {\bar K} }
 }
$
&
$ 0.06 \to 0.21 $  
&
$ 0.48 \pm 0.14 $
\\
$
 { {\Gamma[f_0(1710)\rightarrow \eta\eta']} \over
   {\Gamma[f_0(1710)\rightarrow \eta\eta }
 }
$
&
$ 0.04 \to  0.16 $ 
&
$ < 0.05 $
\\
\hline
\hline
\end{tabular}
\end{center}
\caption[]{Numerical values for the parameters in the 
scalar-pseudoscalar-pseudoscalar Lagrangian [Eq. 
(\ref{L_int_I0})], 
obtained by fitting its 
prediction for decay ratios  (with the specific choice of 
$m[f_0(1370)]=1.35$ 
GeV and $\rho=0$)
to the experimental data 
by WA102 collaboration.} 
\label{T_BBpDDpEF_fit} 
\end{table}

\begin{table}[htbp]
\begin{center}
\begin{tabular}{lll}
\hline
\hline
Decay Widths (MeV) & This Model  & Extracted from other works\\
\hline
$\Gamma[\sigma(550)\rightarrow\pi\pi]$ & $ 53 \to 61 $ & $\approx$ 
100 
\cite{San}\\
$\Gamma[f_0(980)\rightarrow\pi\pi]$ & $107 \to 116$ & $\approx 65$ 
\cite{San}\\
$\Gamma[f_0(1370)\rightarrow\pi\pi]$ &  $121 \to 238$ & $ 34\rightarrow 
175 
$ \cite{Bugg96}\\
$\Gamma[f_0(1370)\rightarrow K{\bar K}]$ & $36 \to 85 $ & $44 
\rightarrow 
240 $ \cite{Bugg96}\\
$\Gamma[f_0(1370)\rightarrow\eta\eta]$ & $< 20 $ & --\\
$\Gamma[f_0(1500)\rightarrow\pi\pi]$ & $103 \to 329 $ & 
$36\rightarrow  65$ \cite{Bugg96}\\
$\Gamma[f_0(1500)\rightarrow K{\bar K}]$ & $2 \to 17$ & $2.4\rightarrow 
7.5$ \cite{Bugg96} \\ 
$\Gamma[f_0(1500)\rightarrow\eta\eta]$ & $< 62 $ & -- 
\\
$\Gamma[f_0(1500)\rightarrow\eta\eta']$ &  $< 23 $ & large 
\cite{Alde88} 
\\
$\Gamma[f_0(1710)\rightarrow\pi\pi]$ &  $< 92$ & $ 1.7\rightarrow 
5.5$ \cite{Longacre86} 
\\
$\Gamma[f_0(1710)\rightarrow K{\bar K}]$ & $3 \to 36$ & $22\rightarrow 
64$ 
\cite{Longacre86}\\ 
$\Gamma[f_0(1710)\rightarrow\eta\eta]$ & $< 8$ & $6\rightarrow 
24$ \cite{Longacre86}\\
$\Gamma[f_0(1710)\rightarrow\eta\eta']$ & $ < 1.5 $ & --\\
\hline
\hline
\end{tabular}
\end{center}
\caption[]{Partial decay widths of $I=0$ scalars predicted by the 
fit in Table 
\ref{T_BBpDDpEF_fit}.} 
\label{T_partial_widths} 
\end{table}

\section{Summary and Conclusion}

In this work we studied the $I=0$ scalar mesons below 2 GeV
[$\sigma (550)$, $f_0(980)$, $f_0(1370), f_0(1500)$, and
$f_0(1710)$] using a non-linear chiral Lagrangian which is
constrained by the mass and the decay properties of the
$I=1/2$ and $I=1$ scalar meson below 2 GeV [$\kappa(900),
K_0^*(1430), a_0(980)$ and $a_0(1450)$]. In this framework
the lowest-lying four-quark scalar meson nonet $N$ mixes
with the next-to-lowest lying two-quark nonet $N'$ and a
scalar glueball $G$.  We showed that this model can
describe the mass spectrum of the scalars, and studied the
correlation between the mass of $f_0(1370)$ and the
substructure of these states. We showed that consistency of
this model with the experimental mass spectrum favors $
1.31\hskip .1cm {\rm GeV}\le m[f_0(1370)] \le 1.45$ GeV,
and sets a bound on the scalar glueball mass in the 1.47
GeV to 1.64 GeV range. We also showed that it is important
to take into account the coupling of the scalar glueball to
$N$ as well as to $N'$.  We found that the $\sigma(550)$ is
mainly a non-strange four quark state, whereas the
substructure of other $I=0$ states is sensitive to the mass
of the $f_0(1370)$. The numerical results show that the
$f_0(1500)$ and $f_0(1710)$ have significant
glueball admixtures.  We also investigated the interaction
Lagrangian and gave a preliminary study of the decay widths
of the $I=0$ scalars into various pseudoscalar-pseudoscalar
channels. Probing scalar-pseudoscalar-pseudoscalar
couplings is important for low energy processes such as
$\eta'\rightarrow 3 \pi$, $\eta'\rightarrow\eta\pi\pi$, in
which the scalar mesons are expected to play important
roles \cite{Far,e3p}, and therefore are interesting
directions for future works. It is also interesting to
examine this model with higher order effects, such as
higher derivative terms or more complex mixing terms
between two and four quark nonets.

\acknowledgments

The author wishes to thank A. Abdel-Rahim, D. Black and J.
Schechter for very helpful discussions.  This work has been
supported by 2003 grant from the State of New York/UUP
Professional Development Committee; 2003 grant from the
joint Labor-Management Institutional Development Awards
Committee; and 2003 Grant from the School of Arts and
Sciences, SUNY Institute of Technology.

\begin{appendix}
\section{The rotation matrices}  

\subsection{$m[f_0(1370)]=1.35 {\rm GeV}$ and $\rho$ a variable:}

\begin{equation}
K^{-1}_{(\rho=-0.5 {\rm GeV}^2)}  =
\left[ \begin{array}{ccccc}
  0.829 &  0.033 &  0.030 &  0.556 &  0.034  \\
 -0.289 &  0.605  &  0.648 &  0.362 & -0.009 \\
 -0.431 & -0.099  & -0.491 &  0.692 & -0.292 \\
 -0.066 &  0.411  & -0.430 &  0.049 &  0.800 \\
  0.197 &  0.674  & -0.393 & -0.280 & -0.524 
\end{array}
\right]
\label{mass_matrix_rhom5}
\end{equation}

\begin{equation}
K^{-1}_{(\rho=-0.4 {\rm GeV}^2)}  =
\left[ \begin{array}{ccccc}
  0.884 & -0.047 & -0.047 &  0.463 & -0.003  \\
 -0.224 &  0.402 &  0.685 &  0.535 & -0.183  \\
 -0.368 & -0.362 & -0.482 &  0.616 & -0.349  \\
 -0.128 &  0.602 & -0.445 &  0.266 &  0.594  \\
  0.129 &  0.586 & -0.315 & -0.223 & -0.701
\end{array}
\right]
\label{mass_matrix_rhom4}
\end{equation}

\begin{equation}
K^{-1}_{(\rho=-0.2 {\rm GeV}^2)}  =
\left[ \begin{array}{ccccc}
  0.927 & -0.138 & -0.292  & 0.155  & 0.108  \\
  0.039 & -0.107 &  0.429  & 0.783  & -0.436  \\
  0.329 & 0.342  & 0.804   & -0.293 & 0.210  \\
 -0.141 & 0.455  & -0.126  & 0.524  & 0.694  \\
  0.104 &  0.803 & -0.264  & -0.042 & -0.522
\end{array}
\right]
\label{mass_matrix_rhom2}
\end{equation}

\begin{equation}
K^{-1}_{(\rho=0.075 {\rm GeV}^2)}  =
\left[ \begin{array}{ccccc}
  0.842 & -0.016 & -0.490 & -0.183 &  0.133  \\
  0.274 & -0.531 &  0.156 &  0.769 & -0.167  \\
  0.414 &  0.062 &  0.852 & -0.232 &  0.212  \\
 -0.077 &  0.302 & -0.075 &  0.435 &  0.842  \\
  0.199 &  0.789 &  0.057 &  0.365 & -0.448  
\end{array}
\right]
\label{mass_matrix_rhom075}
\end{equation}

\subsection{$\rho=0$ and $m[f_0(1370)]$ a variable:}

\begin{equation}
K^{-1}_{(m[f_0(1370)]=1.34 {\rm GeV})}  =
\left[ \begin{array}{ccccc}
  0.844 & -0.003 & -0.490 & -0.159 & 0.149   \\
  0.298 & -0.506 &  0.220 &  0.762 & -0.163  \\
  0.398 &  0.348 &  0.798 & -0.096 &  0.274  \\
 -0.129 &  0.558 & -0.273 &  0.602 &  0.485  \\
 -0.155 & -0.559 &  0.028 & -0.148 &  0.801
\end{array}
\right]
\label{mass_matrix_mf134}
\end{equation}

\begin{equation}
K^{-1}_{(m[f_0(1370)]=1.36 {\rm GeV})}  =
\left[ \begin{array}{ccccc}
  0.853 & -0.034 & -0.476 & -0.142  & 0.159  \\
  0.277 & -0.467 &  0.230 &  0.784  & -0.196  \\
  0.391 &  0.267 &  0.821 & -0.149  & 0.283  \\
 -0.126 &  0.486 & -0.215 &  0.554  & 0.629  \\
  0.167 &  0.688 & -0.035 &  0.192  & -0.679
\end{array}
\right]
\label{mass_matrix_mf136}
\end{equation}

\begin{equation}
K^{-1}_{(m[f_0(1370)]=1.38 {\rm GeV})}  =
\left[ \begin{array}{ccccc}
  0.867 & -0.030 & -0.447 & -0.124  &  0.181 \\
  0.241 & -0.481 &  0.180 &  0.785  & -0.251 \\
  0.389 &  0.204 &  0.862 & -0.122  & 0.220  \\
 -0.110 &  0.490 & -0.148 &  0.570  & 0.633  \\
  0.167 &  0.698 & -0.044 &  0.170  & -0.674
\end{array}
\right]
\label{mass_matrix_mf138}
\end{equation}

\begin{equation}
K^{-1}_{(m[f_0(1370)]=1.40 {\rm GeV})}  =
\left[ \begin{array}{ccccc}
  0.883 & -0.031 & -0.417 & -0.100 &  0.189  \\
  0.197 & -0.483 &  0.136 & 0.793  & -0.283  \\
  0.384 &  0.132 &  0.895 & -0.120 &  0.138  \\
 -0.085 &  0.502 & -0.061 &  0.567 &  0.645  \\
  0.163 &  0.705 & -0.049 &  0.158 & -0.671
\end{array}
\right]
\label{mass_matrix_mf140}
\end{equation}

\section{The coupling constants and decay widths}

The coupling of $I=0$ states $f_0^{NS}, f_0^S, {f'}_0^S, {f'}_0^{NS}$ and 
$G$ to different two-pseudoscalar channels are: 

\begin{eqnarray}	
\gamma^{S}_{\pi\pi} &=& - 2 B 
                       \\
\gamma^{NS}_{\pi\pi} &=& -\sqrt{2} (B - A)  
                       \\
\gamma^{S}_{KK} &=& - 2 ( 2 B - A)
                       \\
\gamma^{NS}_{KK} &=& - 2 \sqrt{2} B   
                      \\
\gamma^{S}_{\eta\eta} &=& 
- 
\sqrt{2} (B + D) + {1\over 2} (C + 2 A + 4 D) \, {\rm sin}\, 2\theta_p
- \sqrt{2} (C + D) \, {\rm cos}^2 \, \theta_p
                       \\
\gamma^{NS}_{\eta\eta} &=& 
- 
(B + D) + {1\over \sqrt{2}} (C + 2 D) \, {\rm sin} \, 2\theta_p
- (A + D) \, {\rm cos}^2 \, \theta_p - C \, {\rm sin}^2 \, \theta_p
                       \\
\gamma^{S}_{\eta\eta'} &=& 
- 
\sqrt{2} (C + D) \, {\rm sin}\, 2\theta_p - (C + 2 A + 4 D) \, 
{\rm cos} \, 2\theta_p
                       \\
\gamma^{NS}_{\eta\eta'} &=& 
- 
\sqrt{2} (C + 2 D) \, {\rm cos} \, 2\theta_p
- (A - C + D)\, {\rm sin}\, 2\theta_p 
                      \\
\gamma^{S}_{\eta'\eta'} &=& 
- 
\sqrt{2} (B + D) - {1\over 2} (C + 2 A + 4 D) \, {\rm sin}\, 2\theta_p
- \sqrt{2} (C + D)\, {\rm sin}^2\, \theta_p
                       \\
\gamma^{NS}_{\eta'\eta'} &=& 
- 
(B + D) - {1\over \sqrt{2}} (C + 2 D) \, {\rm sin} \, 2\theta_p
- (A + D)\, {\rm sin}^2 \, \theta_p - C \, {\rm cos}^2 \, \theta_p
                       \\
\gamma'^{S}_{\pi\pi} &=& -\sqrt{2} (B' - A')  
                       \\
\gamma'^{NS}_{\pi\pi} &=& - 2 B' 
                       \\
\gamma'^{S}_{KK} &=& - 2 \sqrt{2} B'   
                      \\
\gamma'^{NS}_{KK} &=& - 2 ( 2 B' - A')
                       \\
\gamma'^{S}_{\eta\eta} &=& 
- 
(B' + D') + {1\over \sqrt{2}} (C' + 2 D') \, {\rm sin} \, 2\theta_p
- (A' + D') \, {\rm cos}^2 \, \theta_p - C' \, {\rm sin}^2 \,  \theta_p
                       \\
\gamma'^{NS}_{\eta\eta} &=& 
- 
\sqrt{2} (B' + D') + {1\over 2} (C' + 2 A' + 4 D') \, {\rm sin}\, 
2\theta_p
- \sqrt{2} (C' + D') \, {\rm cos}^2 \, \theta_p
                       \\
\gamma'^{S}_{\eta\eta'} &=& 
- 
\sqrt{2} (C' + 2 D') \, {\rm cos} \, 2\theta_p
- (A' - C' + D')\, {\rm sin}\, 2\theta_p 
                       \\
\gamma'^{NS}_{\eta\eta'} &=& 
- 
\sqrt{2} (C' + D') \, {\rm sin}\, 2\theta_p - (C' + 2 A' + 4 D') \, 
{\rm cos} \, 2\theta_p
                       \\
\gamma'^{S}_{\eta'\eta'} &=& 
- 
(B' + D') - {1\over \sqrt{2}} (C' + 2 D') \, {\rm sin} \, 2\theta_p
- (A' + D')\, {\rm sin}^2 \, \theta_p - C' \, {\rm cos}^2 \,  \theta_p
                       \\
\gamma'^{NS}_{\eta'\eta'} &=& 
- 
\sqrt{2} (B' + D') - {1\over 2} (C' + 2 A' + 4 D') \, {\rm sin}\, 
2\theta_p
- \sqrt{2} (C' + D')\, {\rm sin}^2\, \theta_p
                       \\
\gamma^G_{\pi\pi} &=& - \sqrt{2} E 
                       \\
\gamma^G_{K K} &=& - 2 \sqrt{2} E 
                       \\
\gamma^G_{\eta\eta} &=& -  E - F (1 + {\rm cos}^2 \, \theta_p - \sqrt{2}\,
{\rm sin} \, 2 \theta_p )
                       \\
\gamma^G_{\eta\eta'} &=&  - F ({\rm sin} \, 2 \theta_p 
+ 2 \sqrt{2}\,  {\rm cos} \, 2\theta_p)
                       \\
\gamma^G_{\eta'\eta'} &=& -  E - F (1 + {\rm sin}^2 \, \theta_p + 
\sqrt{2}\,
{\rm sin} \, 2 \theta_p )
 \end{eqnarray} where $\theta_p$ is the pseudoscalar mixing angle defined
as 
\begin{equation}
\left( 
\begin{array}{c} 
\eta \\ 
\eta'
\end{array}
\right)
=
\left(
\begin{array}{cc}
{\rm cos} \hskip .1cm \theta_p & -{\rm sin}  \hskip .1cm \theta_p \\
{\rm sin} \hskip .1cm \theta_p & {\rm cos}  \hskip .1cm \theta_p 
\end{array}
\right)
\left( 
\begin{array}{c} 
(\phi_1^1 + \phi_2^2)/ \sqrt{2} \\ 
\phi_3^3
\end{array}
\right)
\end{equation}
with $\theta_p\approx 37^o$ \cite{MS77}.

The two-body partial decay widths of physical states ${\bf F}_i$ are:

\begin{eqnarray}
\Gamma[ {\bf F}_i \to \pi\pi] &=& 
3 \left(
          { 
              { q  {\gamma^i_{ \pi\pi }}^2  } 
                      \over 
              {32 \pi {M_{{\bf F}_i}}^2}
          }
   \right) \left[  { M_{{\bf F}_i}^2 - 2 m_\pi^2}
           \right]^2 \\
\Gamma[ {\bf F}_i \to K {\bar K}] &=& 
 \left(
          { 
              { q  {\gamma^i_{ K K }}^2  } 
                      \over 
              {32 \pi {M_{{\bf F}_i}}^2}
          }
   \right) \left[  { M_{{\bf F}_i}^2 - 2 m_K^2}
           \right]^2 \\
\Gamma[ {\bf F}_i \to \eta\eta] &=& 
2 \left(
          { 
              { q  {\gamma^i_{ \eta\eta }}^2  } 
                      \over 
              {32 \pi {M_{{\bf F}_i}}^2}
          }
   \right) \left[  { M_{{\bf F}_i}^2 - 2 m_\eta^2}
           \right]^2 \\
\Gamma[ {\bf F}_i \to \eta\eta'] &=& 
\left(
          { 
              { q  {\gamma^i_{ \eta\eta' }}^2  } 
                      \over 
              {32 \pi {M_{{\bf F}_i}}^2}
          }
   \right) \left[  { M_{{\bf F}_i}^2 - (m_\eta^2+m_{\eta'}^2}
           \right]^2 
\end{eqnarray}
where $q$ is the center of mass momentum of the final state mesons,  
and for a general two-body decay $A\to B C$ is given as:
\begin{equation}
q = 
{
        \sqrt{ \left[ m_A^2 - (m_B + m_C)^2\right]
               \left[m_A^2 - (m_B - m_C)^2\right]
             }
                        \over
        {2 m_A}
}
\end{equation}

\end{appendix}

\end{document}